\documentclass[twocolumn,superscriptaddress,groupedaddress,pre]{revtex4}  

\usepackage{graphicx}  
\usepackage{dcolumn}   
\usepackage{bm}        
\usepackage{amssymb}   
\usepackage{color}
\usepackage{setspace}
\usepackage{amsmath}
\usepackage{ulem}

\begin{document}

\title{Information Content of Turbulence}

\author{R.T. Cerbus}
\email{rtc17@pitt.edu}
\affiliation{Department of Physics and Astronomy, University of Pittsburgh, 3941 O'Hara Street, Pittsburgh PA 15221}
\author{W.I. Goldburg}
\affiliation{Department of Physics and Astronomy, University of Pittsburgh, 3941 O'Hara Street, Pittsburgh PA 15221}

\begin{abstract}

We treat a turbulent velocity field as a message in the same way as a book or a picture. All messages can be described by their entropy per symbol $h$, defined as in Shannon's theory of communication. In a turbulent flow, as the Reynolds number $Re$ increases, more correlated degrees of freedom are excited and participate in the turbulent cascade. Experiments in a turbulent soap film suggest that the spatial entropy density $h$ is a decreasing function of $Re$, namely $h \propto -\log Re$ + const. In the logistic map, also analyzed here, increasing the control parameter $r$ increases $h$. A modified logistic map with additional coupling to past iterations suggests the significance of correlations.

\end{abstract}

\maketitle

\section{Introduction}

Any physical system is an information channel, since it is ``communicating its past to its future through its present" \cite{crutchfield2012}. A series of data measured for such a system is thus a message, a sequence of symbols. Information theory is the natural framework for the quantitative study of messages \cite{shannon1964,cover1938,brillouin1962,crutchfield2012}. The entropy $H$, also called the information, plays a central role in the theory. It is a measure of uncertainty or disorder. Shannon's theory of communication has found wide application in genetics \cite{yockey2010}, dynamical systems \cite{eckmann1985, shaw1981} and a variety of other fields in physics \cite{mezard2009}. It has also been used extensively in statistical inference problems \cite{jaynes1982} and in this light has provided an interesting way of interpreting the maximization of entropy in statistical mechanics \cite{jaynes1957}. In this work the entropy density $h$, the entropy per symbol, is used as a measure of the information content in a 2D turbulent flow \cite{kellay2002,boffetta2012}.

When a physical system is probed, it reports to the experimenter an ordered sequence of signals $(s_1, s_2,s_3,...)$ In the present experiments, the measured signal is a spatial sequence of velocities $(u_1, u_2, u_3,...)$ in a turbulent 2D flow. They fluctuate in magnitude about a mean flow speed.

Presumably, the disorder $h$ of fluid flow is relatively small if the flow is almost laminar. In this limit of small Reynolds number $Re$, one expects $h$ to increase with $Re$. In the opposite limit of large $Re$, a so-called inertial range of correlated eddies of various sizes develops. Increased correlations implies added constraints or redundancies, which always decrease the uncertainty and information content of any message \cite{cover1938}. One therefore expects that after passing through a maximum, $h$ will decrease with increasing $Re$.

Turbulence is both a temporal and spatial phenomenon. The fundamental work of Kolmogorov treated the spatial structure of turbulence only \cite{kolmogorov1941,davidson2004}. The work of Kraichnan and others suggest that the spatial features are of primary importance \cite{kraichnan1994, shraiman2000, falkovich2001}. Thus, the expectation that $h$ decreases ought to be true for a spatial series but may not be true for time a series.

The present experiments probe a turbulent system at high $Re$, where indeed $h$ is seen to decrease with $Re$ for a spatial velocity sequence. The near-laminar regime, where one expects $h$ to increase with $Re$, is not experimentally accessible.

Treating a physical system as a source of information has its roots in the early development of nonlinear dynamics and chaos \cite{shaw1981}. Pesin's theorem, which for many chaotic systems equates the sum of positive Lyapunov exponents $\lambda$ with the Kolmogorov-Sinai entropy $h_{KS}$ \cite{eckmann1985}, is a nice example of the connection between chaotic dynamics and information theory. If initial conditions separate exponentially, then as two initially almost indistinguishable trajectories separate, new details are uncovered. If $\lambda$ is large, new information is revealed faster.

The methods and ideas of dynamical systems have been applied to turbulence from the earliest stages \cite{eckmann1985}. In many cases turbulence was the original motivation \cite{brandstater1983,swinney1986}. It has been shown by several novel experiments that the onset of turbulence in many systems can be described by a low-dimensional strange attractor, thus solving a long-standing riddle about the transition from laminar flow \cite{eckmann1985,brandstater1983,swinney1986}. (The onset of turbulence in pipe flow is a murkier subject \cite{eckhardt2008}.) In the case of Taylor-Couette flow \cite{brandstater1983}, $h$ and the largest $\lambda$ increase with the Reynolds number $Re$, which can be thought of as a measure of the nonlinearity or strength of the flow. The usual expectation is that this trend continues as $Re$ increases as suggested by some models and analytic work \cite{yamada1988,ruelle1982}.

The motivation for this study is twofold. Firstly, this seems to be the first study of the spatial disorder of a turbulent velocity field as a function of $Re$. By characterizing the flow with the entropy density $h$, the fundamental role of the cascade in producing correlations is clearly manifested. Secondly, $h$ is one of several fundamental quantities necessary to describe how a system creates and communicates information \cite{crutchfield2012,crutchfield2009}. Uncovering how a system produces, stores and transfers information, should prove to be useful. It provides a new and interesting description of nature but has not yet been applied to many physical systems  \cite{crutchfield2012}.

In order to treat a physical system as a message, the experimental data must be converted to symbols \cite{daw2003}. A partition is defined which separates the data into disjoint slices of size $\epsilon$. Then the data values in each specified range (slice) are assigned to a unique symbol \cite{daw2003, schurmann1996}. That is, if data points are $\epsilon$ apart, they correspond to different symbols. (In some sense all experiments do this because of their limited precision.) The size and location of the divisions can be chosen to faithfully represent the original system even for seemingly coarse partitions \cite{daw2003}. 

Correctly identifying those partitions which completely describe the system (called generating) can be extremely difficult \cite{schurmann1996}. However, much can and has been learned about complex systems such as the brain or turbulence even after converting a data series into a simple binary alphabet \cite{daw2003,palmer2000,lehrman1997,lehrman2001}. Approximate treatments are usually necessary and often useful, as long as they still represent the underlying system \cite{daw2003}. For a chaotic time series, the entropy rate $h$ may approach a constant value ($h_{KS}$) as $\epsilon$ decreases \cite{ott2002,schurmann1996}. For a spatially extended system, one may expect the same.

\section{Entropy and Entropy Estimation}

The entropy of a message is usually defined as \cite{shannon1964}
\begin{equation}
H \equiv - \sum_i p_i \log p_i
\label{eqn:shannon_entropy}
\end{equation}
where $p_i$ is the probability of the $i$th symbol occurring in the message. The natural logarithm is used, giving the entropy in ``nats". One may consider $-\log p_i$ as a measure of the information gained from any one symbol. Thus the entropy is the average information of the message. If the message is completely random, then the surprise and the amount of new information $H$ is maximal. $H$ is generally large for broad distributions \cite{cover1938}. By contrast, a constant, unchanging stream of data will have zero entropy. The message contains no new information and no uncertainty.

However, one must take correlations into account since these always reduce the amount of information a message contains. Consider sequential blocks of symbols of length $L$. The probability of any unique block $x_i^{(L)}$ is $p_i^{(L)}$. The Shannon entropy of single symbols can then be generalized to define the block entropy
\begin{equation}
H_L \equiv - \sum_i p^{(L)}_i \log p^{(L)}_i,
\end{equation}
where the sum is over all blocks $x_i^{(L)}$. This block entropy will diverge as $L$ goes to infinity. Therefore one defines a quantity $h$ \cite{cover1938,crutchfield2003}
\begin{equation}
h = \lim_{L \to +\infty} h_L = \lim_{L \to +\infty} H_{L+1} - H_L.
\label{eqn:entropyrate}
\end{equation}
This $h$ is the extra information one gets from measuring one more symbol. The limit exists for stationary processes \cite{cover1938} and may be reached much sooner than $L = \infty$. In spatially extended systems, such as these turbulence measurements, $h$ is called the entropy density. For a time series, $h$ is called the entropy rate or metric entropy \cite{feldman2002,feldman2003,crutchfield1997,crutchfield2003}. Although this distinction does not affect the analysis, it does influence the interpretation of the turbulence results for $h$. (For instance, there is no Pesin's theorem for the entropy density.) The $h$ estimated here is very different from that considered in previous work where it has been estimated for time series \cite{swinney1986,brandstater1983,yamada1988,ruelle1982}.

\begin{figure}[h!]
\hspace{-1.5em}
\includegraphics[scale = 0.33]{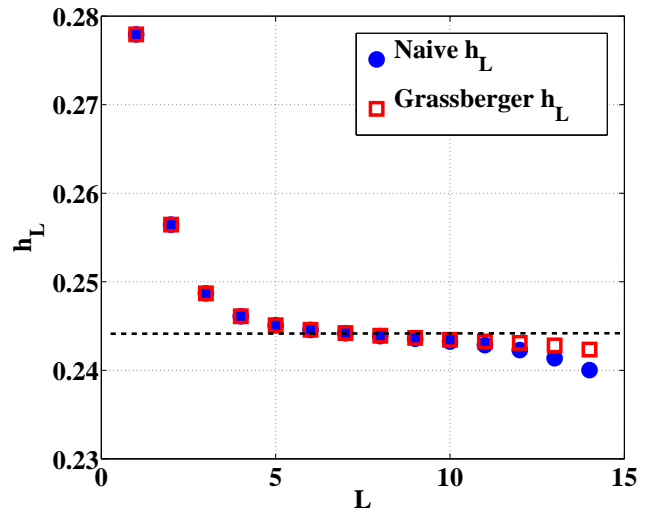}
\caption{$h_L$ as a function $L$ for binarized turbulence data. $h_L$ initially decreases until correlations are no longer important. The value of $h_L$ at the point of inflection is taken to be $h^*$. For larger $L$, $h_L$ decreases again due to undersampling. The naive estimate ($\bullet$) calculates probabilities based on frequency counts. The Grassberger estimate ($\square$) takes the undersampling bias into account \cite{schurmann1996}.}
\label{h_L}
\end{figure}

The above definition already suggests problems one might have in estimating $h$, since the infinite limit is impossible for a finite data set. Fortunately, $h_L$ for real data reaches an asymptote sooner than infinity since correlations are usually finite in scale. Some of the techniques designed to overcome the finite data issues can be found in \cite{schurmann1996}. Most methods involve making an assumption about the distribution of rare events. A technique proposed in \cite{schurmann1996} is used here, although the results are not changed much by its use (see Fig. \ref{h_L}).

In this work the block entropies are used to make an estimate of $h$ by looking for the asymptote of $h_L$ defined by Eq. \ref{eqn:entropyrate} as shown in Fig. \ref{h_L}. In Fig. \ref{h_L}, the naive (frequency count) estimate of $h_L$ is plotted vs. $L$ along with the Grassberger estimate from \cite{schurmann1996}. The dotted line is the value of $h$ determined by the inflection point of $h_L$. The asymptote is usually reached around $L \simeq 10$.

An alternative method for determining a message's information content is based on data compression. The lower limit for the length $S$ to which a message can be compressed from its original length $S_0$, for any compression algorithm, is its entropy: $S \geq H$ \cite{cover1938, schurmann1996, grassberger2002, salomon2007}. Compression algorithms operate by finding redundancies and correlations in data and re-expressing the message in a shorter form. Compression provides a nice way of thinking about how much information is contained in a message, since it reduces the message to its ``essentials". There is no way to shorten a completely random message since each symbol is independent of the others and there can be no compression. For a repetitive stream of symbols (like "...111111...") the message is trivially compressed to almost zero size.

The information content is then \cite{grassberger2002}
\begin{equation}
c = \log D \times {\Big (} \frac{S}{S_0} {\Big )}
\end{equation}
where $D$ is the alphabet size ($e.g.$ for a binary alphabet $D = 2$). The Lempel-Ziv algorithm is optimal in the sense that $c$ converges to $h$ in the limit of infinite $S_0$, so it can be used as another estimate of $h$ and a check on $h^*$. The value of $c$ is independent of file type but does require that the compression program be based on the Lempel-Ziv algorithm. In order to account for the ``overhead" (file headers, etc.), a random data set is compressed and that compression ratio is used to normalize the real data \cite{baronchelli2005}.

Traditionally, $c$ has been given the name of algorithmic or Kolmogorov complexity \cite{cover1938,grassberger2002,baronchelli2005}. This is a measure of the computational complexity of the data set in question. Even if $c$ is not equal to $h$, it is still a measure of the information content of the data \cite{baronchelli2005,kaspar1987,ebeling1997}. It is important to recognize the many limitations involved in calculating information content. At best $h^*$ and $c$ are approximations to $h$, but this does not make them meaningless; they can still be used for comparison \cite{daw2003}.

\section{Results}

\subsection{Logistic Map}

\begin{figure}[h!]
\hspace{-1.5em}
\includegraphics[scale = 0.33]{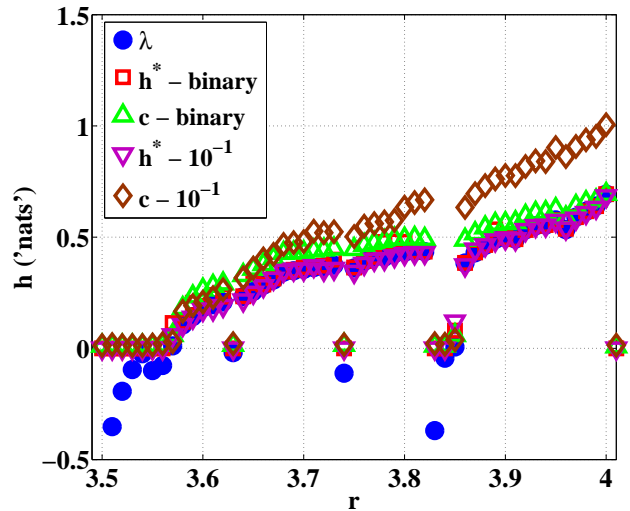}
\caption{Lyapunov exponents $\lambda$ ($\bullet$) for the logistic map plotted as a function of $r$ (see text). The entropy rates were estimated using block entropies $h^*$ (binary: $\square$, $10^{-1}$: $\triangle$) and compression $c$ (binary: $\triangledown$, $10^{-1}$: $\diamond$). Although $h^*$ performs better as an estimate of $h$, all estimates show the same trend.}
\label{logistic_entropy}
\end{figure}

The estimates $h^*$ and $c$ are first applied to the logistic map as a test of the method as well as to illustrate some principles regarding $h$ for chaotic systems. The logistic map is a simple one-dimensional nonlinear map which nicely illustrates chaotic behavior \cite{ott2002}:
\begin{equation}
x_{n+1} = r x_n ( 1 - x_n)
\end{equation}
where $r$ is a parameter that increases the strength of the nonlinearity. As $r$ increases, the system goes through a series of period-doubling bifurcations and eventually becomes chaotic at $r \simeq 3.56995$ ($\lambda > 0$). As usual, $x \in (0,1)$ and $r \in [0,4]$. As mentioned earlier, Pesin's theorem states that the sum of the positive Lyapunov exponents $\lambda$ is equal to $h_{KS}$, as long as the system satisfies certain conditions \cite{eckmann1985}. For the logistic map, which is one-dimensional, there is only one $\lambda$ for each value of $r$. The value of $\lambda$ has been calculated as a function of $r$, using the algorithm in \cite{olivares2008}, and is compared with $h^*$ and $c$ in Fig. \ref{logistic_entropy}. For each value of $r$, a randomly chosen initial condition is iterated $10^6$ times.

Two partitions are shown in Fig. \ref{logistic_entropy}. A binary partition is used where $x = 1$ if  $x \geq 0.5$ and $x = 0$ if $x < 0.5$, so $\epsilon = 0.5$. (The location of this partition divider is important \cite{steuer2001}.) The second partitioning involves simply rounding the data to the first decimal point ($\epsilon = 10^{-1}$) and assigning a symbol to each distinct data value. The estimate $h^*$ performs very well, while $c$ shows significant deviations for the $10^{-1}$ partition. Despite its shortcomings in estimating $h$, $c$ is nonetheless useful as a measure of the information contained in these finite sequences, as discussed above. It follows the same trend as $\lambda$ and reveals the logistic map's information dependence on $r$. The values of $r$ for which $\lambda$ is negative have $h \simeq 0$, since it is positive definite.

Partitions with as few as 2 slices or as many as 1000 slices give essentially the same $h^*$ and $c$. This is because the partitions are all generating, $i.e.$ they represent the dynamics faithfully and the entropy calculated for any of them is the Kolmogorov-Sinai entropy $h_{KS}$ \cite{schurmann1996}. In other words, anything smaller than a binary partition is overkill. This isn't always true, but suggests that crude representations of data can still capture important features. This emboldens us to do the same for turbulence, to be discussed in Section C.

Once the transition to chaos occurs, $\lambda$ and the estimates of $h$ increase almost monotonically. There are several isolated regions where the logistic map returns to periodic behavior \cite{ott2002} and so $\lambda < 0$ and $h \simeq 0$. The general behavior appears to be that as the strength of the nonlinearity increases (see Fig. \ref{logistic_entropy}), so does $h$. Chaos creates information. Similar behavior was observed at the onset of turbulence in Taylor-Couette flow \cite{brandstater1983}. This increase in $h$ for the logistic map is accompanied by a decrease in the strength of correlations, as will be shown shortly. 

\subsection{Modified Logistic Map}

\begin{figure}[h!]
\hspace{-1.5em}
\includegraphics[scale = 0.33]{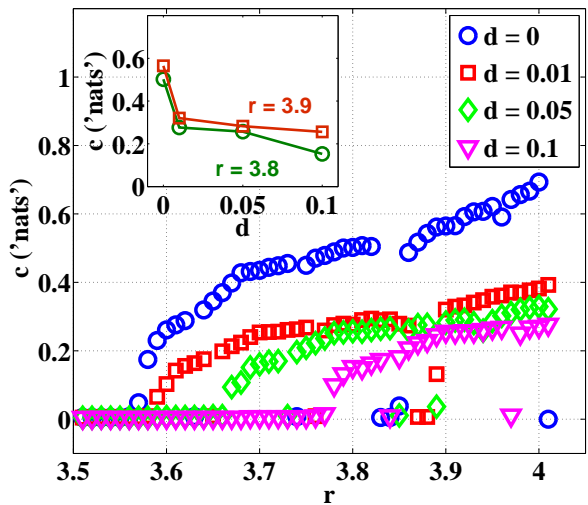}
\caption{Entropy rate estimate $c$ for the modified logistic map using a binary partition. (Recall that a binary partition is generating for the logistic map.) As $d$ increases, $c$ is lowered considerably. ($d$ = 0: $\bigcirc$, $d$ = 0.01: $\square$, $d$ = 0.05: $\diamond$, $d$ = 0.1: $\triangledown$.) Inset: $c$ vs. $d$ for fixed $r$ in the chaotic regime.}
\label{logistic_corr}
\end{figure}

In order to get a better picture of the importance of correlations, a modified logistic map is introduced to explicitly increase correlations through a term that couples to previous iteration values further back than one. Denoting
\begin{equation}
f(x) = rx(1-x),
\end{equation}
the modified logistic map is defined as
\begin{equation}
x_{n+1} = f(x_n) + d {\Big [} \frac{f(x_{n-2}) + f(x_{n-1})}{2} - f(x_n) {\Big ]}
\end{equation}
where $d$ is the coupling strength. This modification is really a kind of logistic delay map \cite{ott2002}. Now using three random intial conditions, this map is also iterated $10^6$ times and the compression estimate is used to compare $h$ for different values of $d$. The results are shown in Fig. \ref{logistic_corr}.

\begin{figure}[h!]
\hspace{-1.5em}
\includegraphics[scale = 0.31]{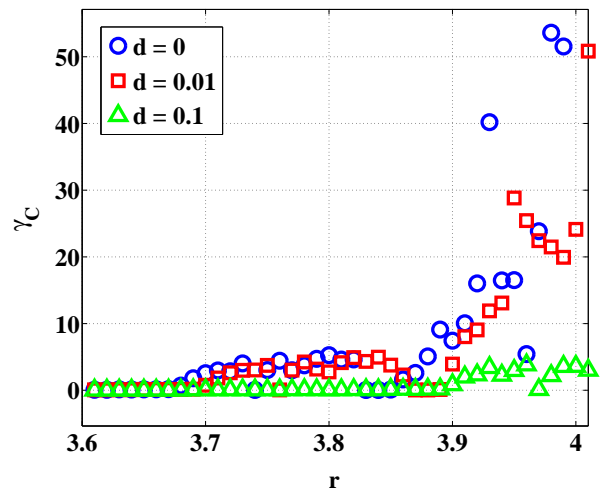}
\caption{Decay rate of mutual information for logistic map and modified logistic map as a function of $r$ using a binary partition ($d$ = 0: $\bigcirc$, $d$ = 0.01: $\square$, $d$ = 0.1: $\triangle$). The decay rate here was calculated as the reciprocal of the area underneath the mutual information curve: $1/\int_0^{\infty} I(\Delta) d\Delta$. If this is large, then the correlations are weak. As the coupling $d$ increases, the correlations get stronger.}
\label{logistic_mutual_info}
\end{figure}

Even for small $d$, $c$ is changed drastically. As $d$ is increased, $c$ is decreased more and the transition to chaos shifts to larger values of $r$. This suggests that in addition to decreasing $h$, correlations can also act to suppress the chaotic transition. 

In order to quantify correlations for messages, it is useful to introduce the mutual information $I$ \cite{cover1938}. This is a measure of the information shared between two variables. For two variables $x$ and $y$ it is
\begin{equation}
I(x;y) \equiv \sum_{x,y} p(x,y) \log \frac{p(x,y)}{p(x) p(y)} = H(x) - H(x|y)
\end{equation}
where $p(x,y)$ is the joint probability of $x$ and $y$ and the second equality shows that the mutual information may also be thought of as the information about variable $x$ minus the information about $x$ given knowledge of $y$. When $x$ and $y$ are uncorrelated, $I(x;y) = 0$. When the two variables $x$ and $y$ are symbols separated by a certain number of symbols $\Delta$ ($x(i)$, $y = x(i+\Delta)$), then $I(\Delta)$ becomes like an autocorrelation function for symbolic sequences \cite{ebeling1995,li1990}. For the logistic map and modified logistic map, $\Delta$ is a temporal interval while for the turbulence measurements $\Delta$ is a spatial interval.

The mutual information is observed to decay exponentially for the chaotic regime of the logistic map and the logistic delay map, with a decay rate $\gamma_C$ that increases with $r$ (see Fig. \ref{logistic_mutual_info}). Put another way $1/\gamma_C$, which can be thought of as a correlation time, decreases with $r$. The correlations are thus decreasing as the strength of the nonlinearity increases, which corresponds well with the understanding that $h$ is reduced by correlations. Figure \ref{logistic_mutual_info} shows $\gamma_C$ as a function of $r$ for three different values of $d$. The addition of coupling has increased the strength of the correlations, and mirrors the drop in $c$.

\subsection{Turbulence}

\begin{figure}[h!]
\hspace{-1.5em}
\includegraphics[scale = 0.37]{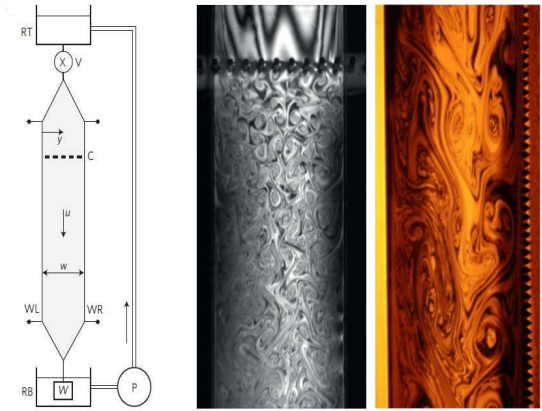}
\caption{Left: Experimental setup showing the reservoirs (RT,RB), pump (P), valve (V), comb (C), blades (WL,WR) and weight (W). Center: Fluctuations in film thickness from turbulent velocity fluctuations for smooth walls and a comb. Right: Fluctuations created by a rough wall. The width $w$ of the channels is several cm.}
\label{setup}
\end{figure}

Now consider the real physical system of a turbulent soap film, which is a good approximation to 2D turbulence since the film is only several $\mu$m thick \cite{kellay2002,boffetta2012}. The soap solution is a mixture of Dawn$^{\small \textregistered}$  (2$\%$) detergent soap and water with 1.5 $\mu$m particles added for laser doppler velocimetry (LDV) measurements. Figure \ref{setup} is a diagram of the experimental setup.

The soap film is suspended between two vertical blades connected to a nozzle above and a weight below by nylon fishing wire. The nozzle is connected by tubes to a valve and a reservoir which is constantly replenished by a pump that brings the soap solution back up after it has flowed through. The flow is gravity-driven. Typical centerline speeds $u$ are several hundred cm/s with rms fluctuations $u'$ ranging from roughly 1 to 30 cm/s. The channel width $w$ is usually several cm. 

Turbulence in the soap film is generated by either (1) inserting a row of rods (comb) perpendicular to the film or (2) replacing one or both smooth walls with rough walls (saw blades), with the comb removed. When protocol (1) is used decaying 2D turbulence results which is almost always accompanied by the direct enstrophy cascade \cite{kellay2002,boffetta2012}. If procedure (2) is used, then forced 2D turbulence can be generated with an inverse energy cascade \cite{kellay2002,boffetta2012}. The ability to see the inverse energy cascade depends sensitively on the flux and channel width. This sensitivity is decreased if two rough blades are used. The type of cascade is determined by measuring the one-dimensional velocity energy spectrum $E(k)$, where $\frac{1}{2}u'^2 = \int_0^{\infty} E(k) dk.$ 

Although a condensate has been observed in some 2D turbulent systems \cite{boffetta2012}, it is not present in this one. A condensate is revealed by a sharp spike in $E(k)$, which is never observed.  In other experimental arrangements, two slopes are seen in a log-log plot of $E(k)$ vs. $k$, indicating a dual cascade of both energy and enstrophy \cite{rutgers1998,boffetta2012,kellay2002}. For these experiments only one slope is observed.

Measurements of the velocity are usually taken near the vertical middle of the channel. In all cases, the data are obtained for the longitudinal velocity component at the horizontal center of the channel. The data rate is $\simeq$ 5000 Hz and the time series typically had more than $10^6$ data points. For this system, the time series should be thought of as a spatial series by virtue of Taylor's frozen turbulence hypothesis \cite{davidson2004,kellay2002,boffetta2012}. Its validity has been thoroughly tested for this system \cite{belmonte2000}. The fact that these measurements involve a spatial series rather than a time series is a crucial point.

\begin{figure}[h!]
\hspace{-1.5em}
\includegraphics[scale = 0.31]{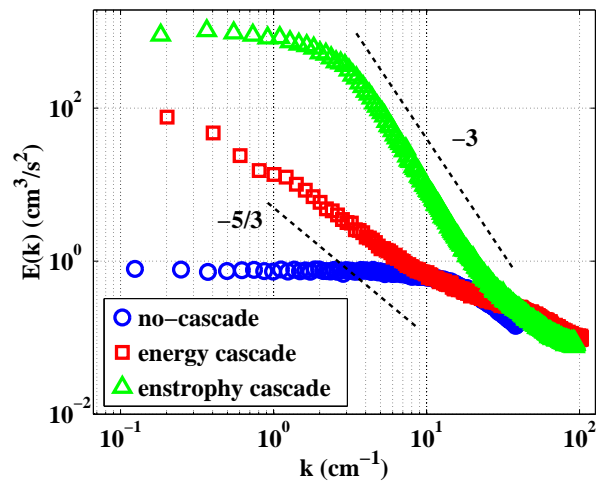}
\caption{Representative one-dimensional energy spectra in a log-log plot of $E(k)$ vs. $k$. The enstrophy cascade ($\triangle$) has a slope close to -3 while the energy cascade ($\square$) has a slope close to -5/3. The flat curve ($\bigcirc$) has no cascade.}
\label{spectra}
\end{figure}

With this high data rate, the smallest turbulent scales are easily resolved. A number of measurements were taken near the top of the channel where the flow is still quite slow. In this case there is no power law scaling in $E(k)$ and so apparently no cascade, although the flow is not laminar ($u' \neq 0$). Some representative spectra are shown in Fig. \ref{spectra}.

Before converting the velocity data into symbols, the mean velocity is subtracted out and the result divided by $u'$. This was done to have a similar alphabet size for different $Re$ and seems a natural way to treat the data. The velocity data were then partitioned in a similar way to the logistic map. That is, the data were separated into slices of various sizes and then converted into symbols. In the turbulence case a binary partition means that a 1 is assigned if the velocity is above the mean value and 0 if below.

\begin{figure}[h!]
\hspace{-1.5em}
\includegraphics[scale = 0.31]{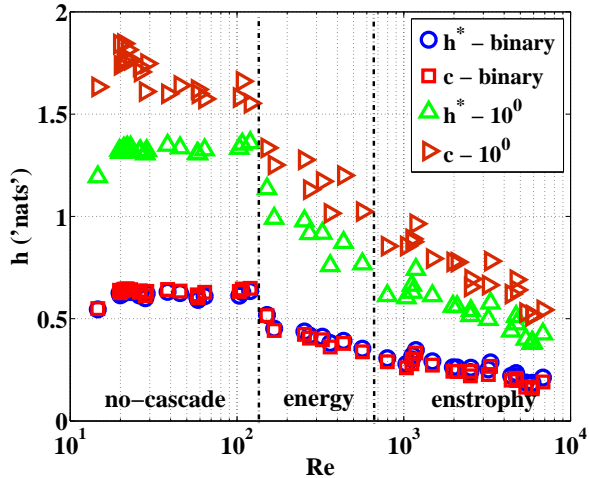}
\caption{Entropy density estimates of $h$ vs. $Re$ for the 2D turbulent data. $h$ is a decreasing function of $Re$. The flat, higher $h$ region corresponds to the no-cascade data. The decay begins with the emergence of the cascade. A binary partition ($h^*$: $\bigcirc$, $c$: $\square$) and a second partition where the data is saved to the nearest integer ($h^*$: $\triangle$, $c$: $\triangleright$) are shown here. Dividing lines are shown that separate the data into no-cascade, energy and enstrophy regions respectively.}
\label{turbulence_entropy}
\end{figure}

The main results of this paper appear in Fig. \ref{turbulence_entropy}, which is a plot of $h^*$ and $c$ vs. $Re$, where $Re \equiv u'w/\nu$ and $\nu$ is the kinematic viscosity. The Reynolds number $Re$ is a measure of the nonlinearity of the system, much like $r$ for the logistic map. Four different estimates of $h$ are shown in Fig. \ref{turbulence_entropy}. The open circles ($\bigcirc$) and squares ($\square$) show $h^*$ and $c$ respectively for the binary partition. The two upper data sets ($\triangle$,$\triangleright$) are $h^*$ and $c$ for a finer partition where the velocity data are distinguished by their first significant figure ($\epsilon = 1$). The same trend is shown by both partitions and for all partitions studied, namely that $h^*$ and $c$ are decreasing functions of $Re$.  

Note that $h^*$ and $c$ are very weakly dependent on $Re$: $h^* \propto -\log Re$ after an initial plateau. The decrease begins as soon as a cascade appears, as seen in Fig. \ref{turbulence_entropy}. The decrease is independent of the type of cascade, as both the energy and enstrophy cascade data are present in the figure. The flat region at low $Re$ corresponds to the data without a cascade.

At first glance this result seems surprising, but the decrease is in accord with the common picture of the turbulent cascade \cite{davidson2004, kellay2002,boffetta2012}. The energy (enstrophy) flows from one scale $r$ to nearby spatial scales. The eddies participating in the cascade are necessarily correlated and the extent of the inertial range (cascade region) increases with $Re$. Since laminar flow is not disordered at all, this implies that $h$ passes through a local maximum at an intermediate value of $Re$. It is regrettable that the soap film is not stable at low $Re$, thus hindering the observation of this local maximum. 

Although the system under study here is two dimensional, the same decrease in $h$ should also hold for three dimensional turbulence in the fully developed regime. It should be noted that the results in Fig. \ref{turbulence_entropy} are somewhat similar to that of Wijesekera {\it et. al.} in their study of spatial density fluctuations in the ocean \cite{wijesekera1997}. However, the behavior of the spatial entropy density $h$ vs. $Re$ observed here is quite different from that of the temporal entropy rate studied previously \cite{yamada1988,ruelle1982,brandstater1983,swinney1986}.

\begin{figure}[h!]
\hspace{-1.5em}
\includegraphics[scale = 0.31]{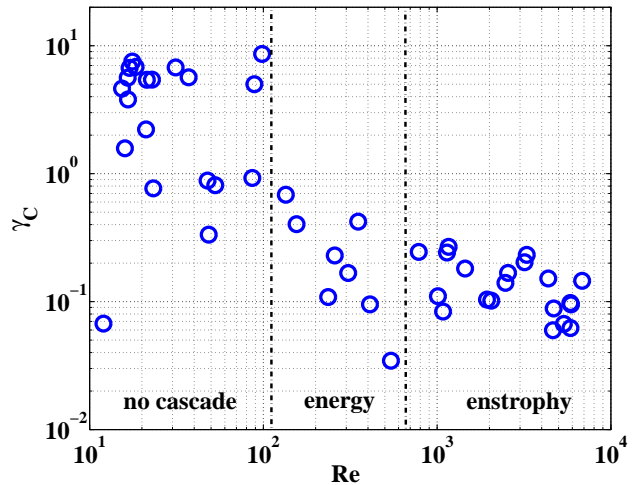}
\caption{Spatial decay rate of mutual information for turbulence data using a partition where data are saved to the nearest integer. The decay rate here was calculated as the reciprocal of the area underneath the mutual information curve: $1/\int_0^{\infty} I(\Delta) d\Delta$. If this is large, then the correlations are weak.  Dividing lines are shown that roughly separate the data into no-cascade, energy and enstrophy regions respectively. As $Re$ increases, the correlations get stronger.}
\label{turbulence_decay}
\end{figure}

The spatial correlations in the flow are becoming increasingly important as $Re$ increases. This is evidenced by the decrease in the (spatial) decay rate $\gamma_C$ for the mutual information $I$($\Delta$) as shown in Fig. \ref{turbulence_decay}. The increased strength of the correlations is responsible for the decrease in $h$, which follows from its definition in Eq. \ref{eqn:entropyrate}.

\begin{figure}[h!]
\hspace{-1.5em}
\includegraphics[scale = 0.33]{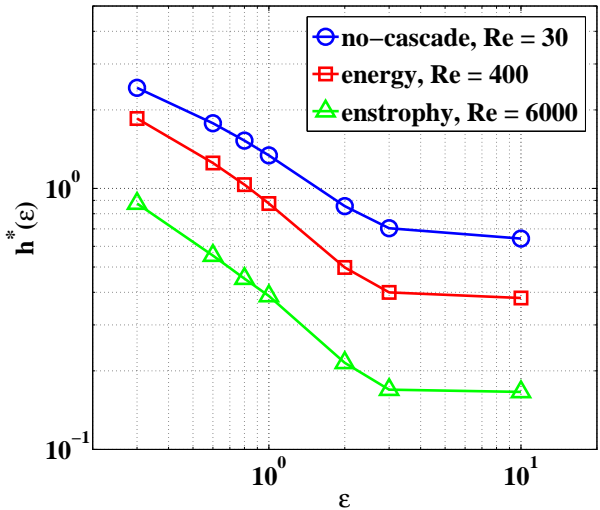}
\caption{Entropy density $h^*$ as a function of the partition size $\epsilon$ for three different $Re$. (The largest partition size corresponds to binarized data.) The three curves correspond to the no-cascade ($\bigcirc$), energy cascade ($\square$) and enstrophy cascade ($\triangle$) data. Despite the significant $\epsilon$-dependence, none of the curves intersect. This means that the $Re$-dependence is $\epsilon$-independent.}
\label{entropy_partition}
\end{figure}

Unlike the logistic map, the turbulence data is more sensitive to the size of the partition $\epsilon$ when converting to symbols. The location of the dividers is important for the coarser partitions \cite{steuer2001}, but as the partition size decreases the results are not sensitive to this placement. It is generally true that $h$ depends on the partition size $\epsilon$ \cite{gaspard1993} as shown in Fig. \ref{entropy_partition}. In Fig. \ref{entropy_partition}, $h^*$ is plotted as a function of $\epsilon$ for three different values of $Re$. Although $h^*$ increases as $\epsilon$ decreases, the curves never cross for the different $Re$. As $\epsilon$ decreases, more detailed information is described by the symbols. Although $c$ is not a reliable estimate of $h$ at the finer partitions, it is still an indicator of the information content of the data streams and also shows the same decrease with $Re$ \cite{baronchelli2005}. The important point is that the general behavior of $c$ and $h^*$ is the same for partitions of all sizes.

Although the selection of a correct partition is trickier for the turbulent data $h^*$ and $c$ are more sensitive to $\epsilon$ for the turbulence data, they show that the spatial disorder decreases with $Re$ at each level of descriptive precision. An estimate can be made for the smallest size of the partition needed to capture the entire inertial range, based on the smallest eddy's characteristic velocity $u_{\eta}$ \cite{davidson2004,gaspard1993,wang1992}. Simple estimates show that the smaller partitions are fine enough to resolve a fluctuation of this magnitude for all $Re$. It is surprising that even a binary partition captures the main features: $h \propto -\log Re$. This suggests that one may fruitfully study turbulence just by looking at the 1s and 0s. Similar studies of complex systems such as the brain and heart and even turbulence \cite{daw2003,lehrman1997,lehrman2001,palmer2000} have also used very coarse partitions.

As an additional test of the validity of this coarse-graining approach, the decay rates calculated from the raw data using the autocorrelation method and with the mutual information were compared for various partitions (data not shown). The $Re$-dependence was almost exactly the same, although there is a shift by a factor of $1/e$ for the mutual information method. Since the entropy is fundamentally connected to correlations, this is strong evidence for the validity of this coarse-graining approach.

Entropy maximization is a familiar principle for solving a variety of problems and is a fundamental principle in equilibrium statistical mechanics \cite{jaynes1957,jaynes1982}. In these problems, understanding the constraints is of paramount importance. In a turbulent system the constraints are correlations that span a wider range of scales as $Re$ increases. However, this begs the question as to why this happens. Perhaps the organization of the cascade is the response to the system's effort to more efficiently transfer energy (enstrophy) between scales.

\section{Conclusion}

Treating turbulence as a message enables one to quantify the information content in the system through the entropy density $h$. Estimates show that $h$, a measure of disorder, is a decreasing function of $Re$ at large $Re$. The cascade reduces spatial randomness by introducing correlations.

Cascades in turbulence are often thought to arise naturally because of the wide separation between the forcing scale and the dissipative scale, as well as because of some essential features of the Navier-Stoke's equation \cite{davidson2004,sreenivasan1995}. However, this may not be the only way of looking at the issue, just as in mechanics one can use either Newton's laws or a variational principle and get the same answer. Perhaps the underlying reason for the cascade can be connected to the decrease of $h$ as $Re$ increases.

\section{Acknowledgements}

The authors are grateful to Mahesh Bandi for introducing us to this topic. Comments from various referees were also very helpful. This work is supported by NSF Grant No. 1044105 and by the Okinawa Institute of Science and Technology (OIST). R.T.C. is supported by a Mellon Fellowship through the University of Pittsburgh.

\end{document}